\documentclass[utf8]{FrontiersinHarvard} 

\usepackage{url,hyperref,lineno,microtype,subcaption}
\usepackage[onehalfspacing]{setspace}
\usepackage{graphicx}
\usepackage{booktabs}   

\def\keyFont{\fontsize{8}{11}\helveticabold }
\def\firstAuthorLast{J. Zhuo {et~al.}} 
\def\Authors{Jiahui Zhuo\,$^{1,*}$, Arantza Oyanguren\,$^{1}$, Álvaro Fernández Casani\,$^{1}$, Luca Fiorini\,$^{1}$, and Valerii Kholoimov\,$^{1,2}$}

\def\Address{  
$^{1}$IFIC (Instituto de Física Corpuscular), University of Valencia- CSIC, Valencia, Spain \\
$^{2}$École Polytechnique Fédérale de Lausanne (EPFL), Laussene, Switzerland. }
\def\corrAuthor{Jiahui Zhuo,  IFIC - C/. Catedrátic José Beltrán Martínez, 2, 46980 Paterna, Valencia}
\def\corrEmail{Jiahui.Zhuo@cern.ch}

\begin{document}
\onecolumn
\firstpage{1}

\title[Energy Efficiency of a GPU-based computing system for HEP experiments]{Energy efficiency of a GPU-based computing system for High Energy Physics experiments} 

\author[\firstAuthorLast ]{\Authors} 
\address{} 
\correspondance{} 

\extraAuth{}

\maketitle

\begin{abstract}
\section{}
In this paper we introduce the energy efficiency as a new metric for evaluating both hardware platforms based on Graphic Processor Units (GPU), and algorithm optimisations at High Energy Physics (HEP) experiments. We develop a method to compute the energy efficiency for the case of the first high level trigger (HLT1) of the LHCb experiment, relating the throughput with GPU specifications such as the number of cores, clock frequency, memory bandwidth and thermal design power. The model can be extended to other HEP experiments to make decisions and reach sustainable computing ecosystems. 


\tiny
 \keyFont{ \section{Keywords:} Energy efficiency, power consumption, Graphic Processor Units (GPUs), computing, throughput,  LHC} 
\end{abstract}

\section{Introduction}

Many computational approaches in High Energy Physics (HEP), particularly at the Large Hadron Collider (LHC) at CERN, rely at the moment on CPU architectures which lack scalability for the future high-luminosity upgrade (HL-LHC), posing significant energy constraints. High-performance heterogeneous architectures are being explored as alternative strategies to handle the large volume of data. In particular, real time analysis for trigger and reconstruction at the LHCb experiment is an example of the new successful GPU implementation.  
A key question when deploying a GPU-based application is how its throughput scales with the hardware. Understanding this scalability reveals whether the workload is bounded, and how efficiently the algorithms exploit the available parallelism. It also allows projections for future GPU generations without running benchmarks on each one.
Equally important, but often overlooked, is the energy efficiency: how many events the application can process per joule of energy consumed. For a large-scale system like the ones used by CERN experiments, energy efficiency directly determines the electricity cost and the cooling infrastructure required to operate it. 

The LHCb experiment at CERN has successfully deployed a GPU-based High-Level Trigger (HLT1) for its Run 3 data-taking period, which began in 2022, described in \cite{CERN-LHCC-2020-006}. It performs real time event reconstruction and selection at 30 MHz proton-proton collision rate, implying a data volume of around 40 terabits (Tb) per second. 
The system is implemented using approximately 500 NVIDIA RTX A5000. Similar approaches are being investigated by ATLAS and CMS experiments (see \cite{Gessinger:2025wjo,ATL-DAQ-PUB-2025-003,CMS:2023gfb,cms2}). 
The performance of a GPU is characterised by a small set of hardware parameters: the number of CUDA cores (the parallel processing units that execute the computations), the clock frequency (the speed at which each core operates), the memory bandwidth (the rate at which data can be transferred between the GPU memory and the cores) and the thermal design power (TDP, the maximum power the GPU is designed to consume). Different GPU models offer different combinations of these parameters, and the throughput of a given application (the number of events executed per second) depends on how effectively it can exploit them.
 
 As the GPU technology advances, computing farms at LHC experiments need hardware upgrades. The GPU market offers a wide range of models with different combinations of the hardware parameters introduced above. Testing each candidate GPU individually by installing it and running the software is time-consuming and expensive. A predictive model for both throughput and power consumption would allow the LHC collaborations to evaluate the scalability and energy efficiency of candidate GPUs using only their specification parameters, before committing to full benchmarks. Such models also deepen the understanding of the relationship between the hardware and the software, revealing which hardware resources the algorithms benefit from most and where further software optimisation would be most effective.
 
\section{Methodology}
The GPU infrastructure used in this study was partially provided by the HIGH-LOW project at IFIC-Valencia (see \cite{High-low}). 
Ten NVIDIA GPUs are selected for this study. Each belongs to an \emph{architecture}, which defines the design of its streaming multiprocessors, cache hierarchy and instruction set. The four architectures represented in this study are Ampere, Ada Lovelace, Hopper and Blackwell, listed in order of release. Each architecture is manufactured on a particular \emph{fabrication process node}, which determines the transistor density and the power efficiency of the chip. The two Ampere GPUs (RTX A5000 and RTX 3090) use the Samsung 8\,nm process. The remaining eight GPUs use the more recent TSMC 4\,nm process: five Ada Lovelace (RTX 4070 Ti Super, RTX 4080 Super, RTX 5000 Ada, RTX 4090, RTX 6000 Ada), two Blackwell (RTX PRO 4000, RTX PRO 6000), and one Hopper (H100 NVL). The hardware parameters published by NVIDIA in the official product datasheets, referred to as the \emph{specification} parameters, are listed for each GPU in Table~\ref{tab:throughput_gpu_specs}. The GPUs cover a wide range of CUDA core counts (8192 to 24064), boost clock frequencies (1695 to 2617\,MHz), memory bandwidths (576 to 3938\,GB/s), and TDP values (145 to 600\,W).

\begin{table}[htb]
    \centering
    \caption{Specification parameters of the ten GPUs used in this study, grouped by architecture. The boost clock, memory bandwidth and TDP values are taken from the official NVIDIA datasheet.}
    \label{tab:throughput_gpu_specs}
\resizebox{\textwidth}{!}{%
    \begin{tabular}{llrrrr}
        \toprule
        GPU & Architecture & CUDA cores & Boost clock & Memory BW & TDP \\
            &              &            & [MHz]       & [GB/s]    & [W] \\
        \midrule
        RTX A5000         & Ampere       &  8192 & 1695 &  768 & 230 \\
        RTX 3090          & Ampere       & 10496 & 1695 &  936 & 390 \\
        \addlinespace
        RTX 4070 Ti Super & Ada Lovelace &  8448 & 2610 &  672 & 285 \\
        RTX 4080 Super    & Ada Lovelace & 10240 & 2550 &  736 & 320 \\
        RTX 5000 Ada      & Ada Lovelace & 12800 & 2550 &  576 & 250 \\
        RTX 4090          & Ada Lovelace & 16384 & 2520 & 1008 & 450 \\
        RTX 6000 Ada      & Ada Lovelace & 18176 & 2505 &  960 & 300 \\
        \addlinespace
        RTX PRO 4000      & Blackwell    &  8960 & 2617 &  672 & 145 \\
        RTX PRO 6000      & Blackwell    & 24064 & 2617 & 1792 & 600 \\
        \addlinespace
        H100 NVL          & Hopper       & 16896 & 1785 & 3938 & 400 \\
        \bottomrule
\end{tabular}}
\end{table}

The throughput and power measurements are performed using the Allen version \texttt{v7r10p1}, running the \texttt{hlt1\_pp\_default} reconstruction sequence, described in \cite{CERN-LHCC-2020-006}. This sequence is the baseline HLT1 configuration used in LHCb during the Run~3 data taking, and includes about $\mathcal{O}(300)$ reconstruction algorithms executed in the first high level trigger. The application provides a large amount of tasks including tracking the trajectories of charged particles, identifying proton-proton collision points, classifying particles as hadrons or muons, and detecting displaced decay vertices of long-lived particles.


The NVIDIA System Management Interface  program (\texttt{nvidia-smi}) is used to monitor the power consumption of the GPUs used in this work. An external device designed for measuring the power distribution across the server rack, an APC Metered Rack PDU ZeroU 2G AP8, was used in ~\cite{High-low} to perform absolute power measurements and to validate the results on the GPUs.   

\section{Results}

\subsection{GPU performance}

\begin{figure}[htb]
    \centering
    \includegraphics[width=0.55\textwidth]{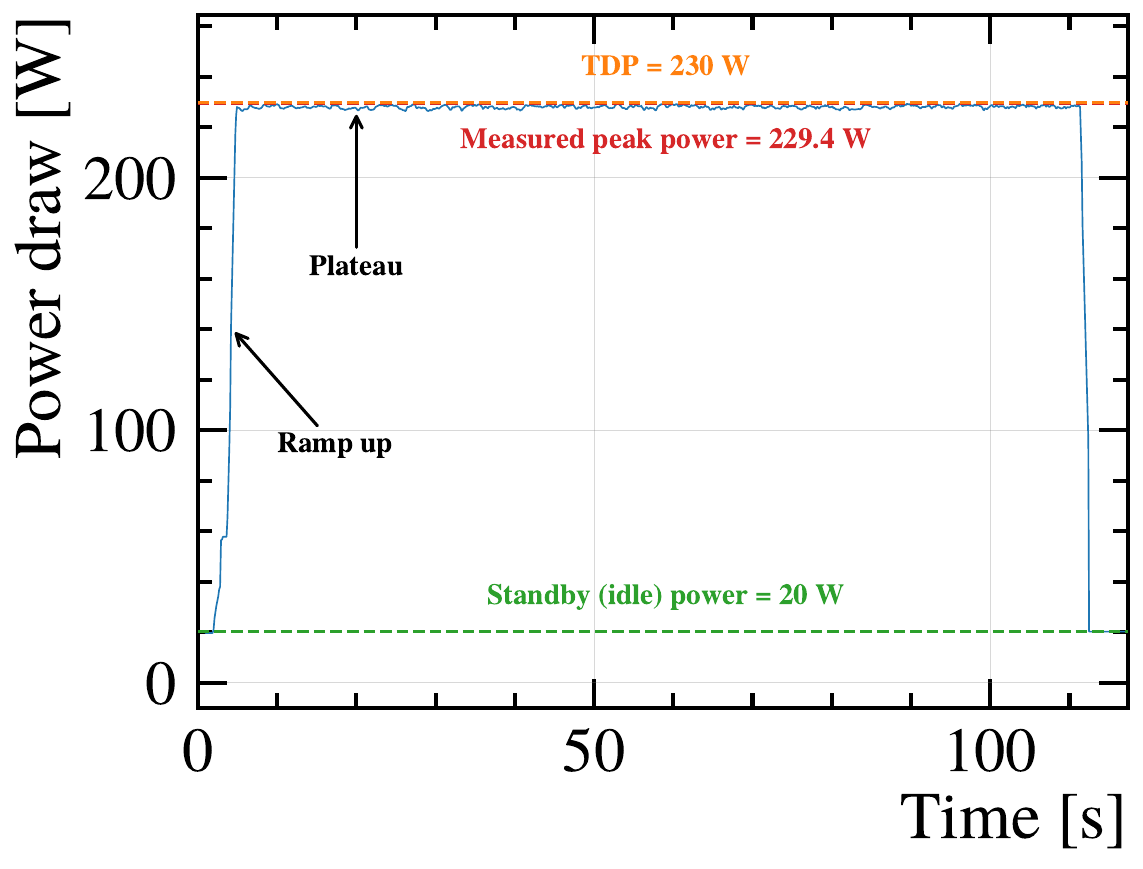}
    \caption{Power draw as a function of time during a single LHCb HLT1 benchmark run of 500k events on the RTX A5000. The measured peak power (229.4\,W) is extracted from the plateau where the GPU utilisation is at 100\%. The TDP (230\,W) and the standby power (20\,W) are shown as dashed lines.}
    \label{fig:throughput_power_measurement}
\end{figure}

During each benchmark run, the GPU is monitored using \texttt{nvidia-smi} at 1\,second intervals. This records the streaming multiprocessor (SM) clock frequency, memory clock frequency and power draw during the benchmark run. Figure~ \ref{fig:throughput_power_measurement} shows the power draw as a function of time during a benchmark run on the RTX A5000. Before the benchmark starts, the GPU draws approximately 20\,W in standby. Once the algorithms begin processing, the power draw ramps up and reaches a stable plateau where the GPU utilisation is at 100\%. The measured quantities (SM clock, memory clock, power draw) are extracted from this plateau, which represents the steady-state operating conditions during continuous data taking.

\begin{figure}[htb]
    \centering
    \includegraphics[width=0.45\textwidth]{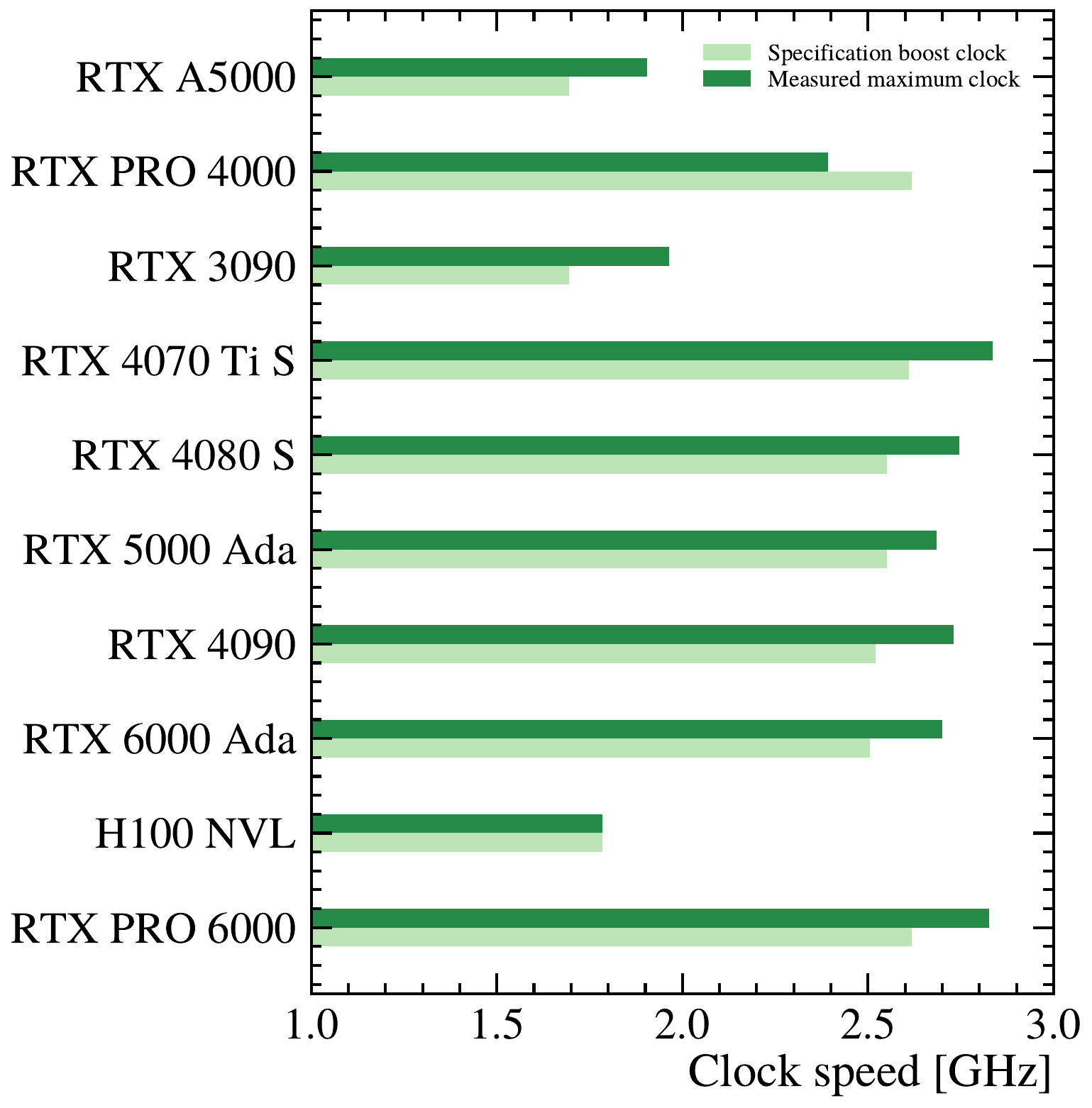}
    \caption{Specification boost clock compared with the measured maximum SM clock during HLT1 execution. Most GPUs exceed their specification boost clock by 5--16\% due to the GPU Boost mechanism.}
    \label{fig:throughput_clock_comparison}
\end{figure}

Figure~\ref{fig:throughput_clock_comparison} compares the measured SM clock frequency from the stable plateau of the power draw curve in Fig.~\ref{fig:throughput_power_measurement} with the specification boost clock for each GPU. Most GPUs run 5--16\% above their specification boost clock. This is expected: the specification boost clock is a guaranteed minimum, not a maximum. Modern NVIDIA GPUs automatically increase their clock frequency when there is enough thermal and power margin, a behaviour known as GPU Boost.

\begin{figure}[htb]
    \centering
    \includegraphics[width=0.45\textwidth]{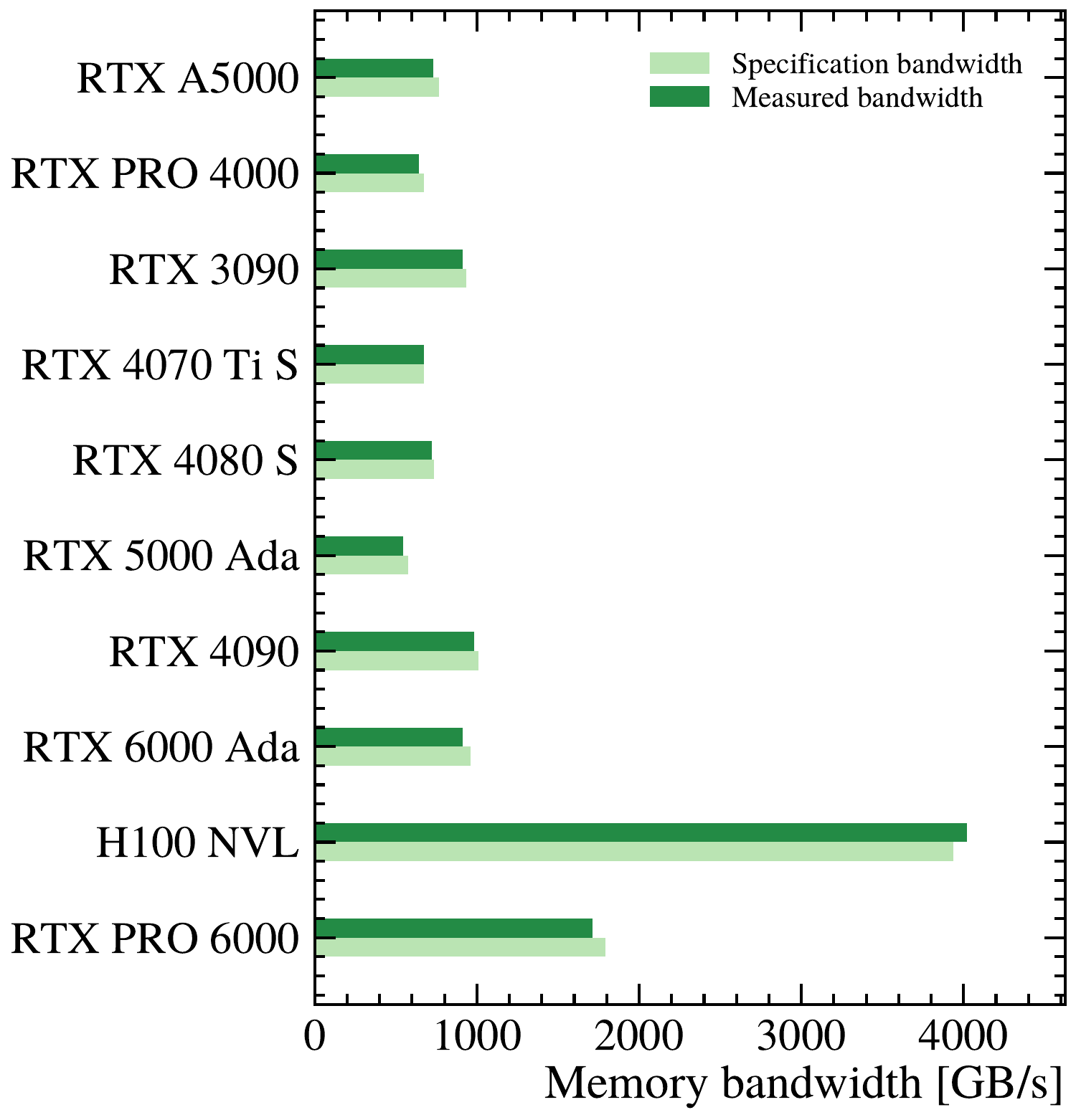}
    \caption{Specification memory bandwidth compared with the measured memory bandwidth during HLT1 execution. The measured bandwidth closely matches the specification for all GPUs.}
\label{fig:throughput_bandwidth_comparison}
\end{figure}

From the measured memory clock frequency, the measured memory bandwidth can be computed as $\text{BW} = 2 \times f_\text{mem} \times W_\text{bus}$, where $f_\text{mem}$ is the measured memory clock and $W_\text{bus}$ is the memory bus width. Figure~\ref{fig:throughput_bandwidth_comparison} compares the result with the specification bandwidth. Unlike the SM clock, the measured memory bandwidth closely matches the specification for all GPUs, because the memory clock does not have a boost mechanism.

\begin{figure}[htb]
    \centering
    \includegraphics[width=0.45\textwidth]{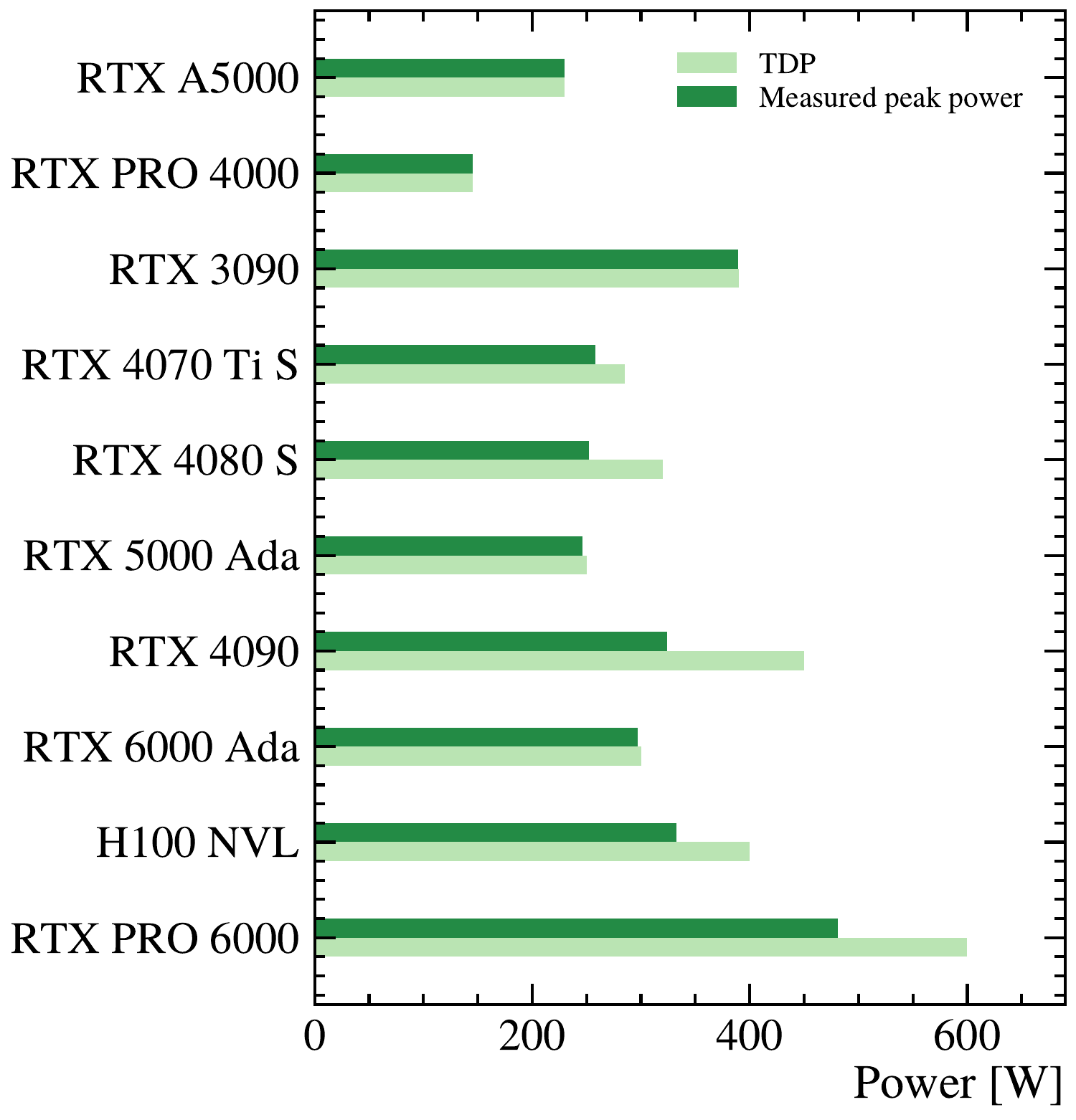}
    \caption{TDP compared with the measured peak power during HLT1 execution. Power-limited GPUs (RTX A5000, RTX 3090, RTX 5000 Ada, RTX 6000 Ada, RTX PRO 4000) reach close to their TDP, while the remaining GPUs run well below it.}
    \label{fig:throughput_power_comparison}
\end{figure}

Figure~\ref{fig:throughput_power_comparison} compares the measured peak power with the TDP for each GPU. Two distinct behaviours are observed. Some GPUs reach very close to their TDP (99--100\%), meaning the HLT1 workload consumes the entire power budget. These GPUs are referred to as \emph{power-limited}. Other GPUs run well below their TDP (72--91\%), meaning the power budget is not the bottleneck. This distinction between power-limited and non-power-limited GPUs is central to the power consumption model developed later.

The comparison in the previous figures show that the actual SM clock frequency and memory bandwidth of a GPU during HLT1 execution do not exactly match the specification values. The SM clock is typically 5--16\% higher than the specification due to GPU Boost, while the memory bandwidth is close to the specification.

\subsection{Throughput model}
\label{sec:throughput_power}

The HLT1 throughput is modelled as a power law of two hardware parameters and fitted to measured values. The first is the total compute capacity, defined as the product of the number of CUDA cores $N_\text{cores}$ and the SM clock frequency $f_\text{clk}$. This product represents the total number of operations per second available on the GPU. The second is the memory bandwidth BW, which determines how fast data can be moved between the GPU memory and the compute units. These two quantities represent the two fundamental resources of any GPU workload: computation and data transfer. The model expresses the throughput TP (in kHz) as
\begin{equation}
    \text{TP} = k \times (N_\text{cores} \times f_\text{clk})^{a} \times \text{BW}^{b} \,,
    \label{eq:throughput_model} 
\end{equation}
where $f_\text{clk}$ is expressed in GHz, BW is expressed in GB/s, $k$ is a normalisation constant, $a$ is the exponent on the compute capacity $N_\text{cores} \times f_\text{clk}$, $b$ is the exponent on the memory bandwidth, and all three are free parameters determined by fitting to the measured throughput of all ten GPUs. A power law is used because GPU applications generally scale less than linearly with hardware resources. For the compute capacity, synchronisation overhead between threads and the fact that not all parts of the workload can run in parallel lead to $a < 1$. For the memory bandwidth, GPUs have multiple levels of cache (L1 and L2), so many memory accesses are served by the cache and do not use the global memory bandwidth. The throughput therefore depends on the bandwidth less strongly than it would without caching, leading to $b < 1$.

\begin{figure}[htb]
    \centering
    \includegraphics[width=0.65\textwidth]{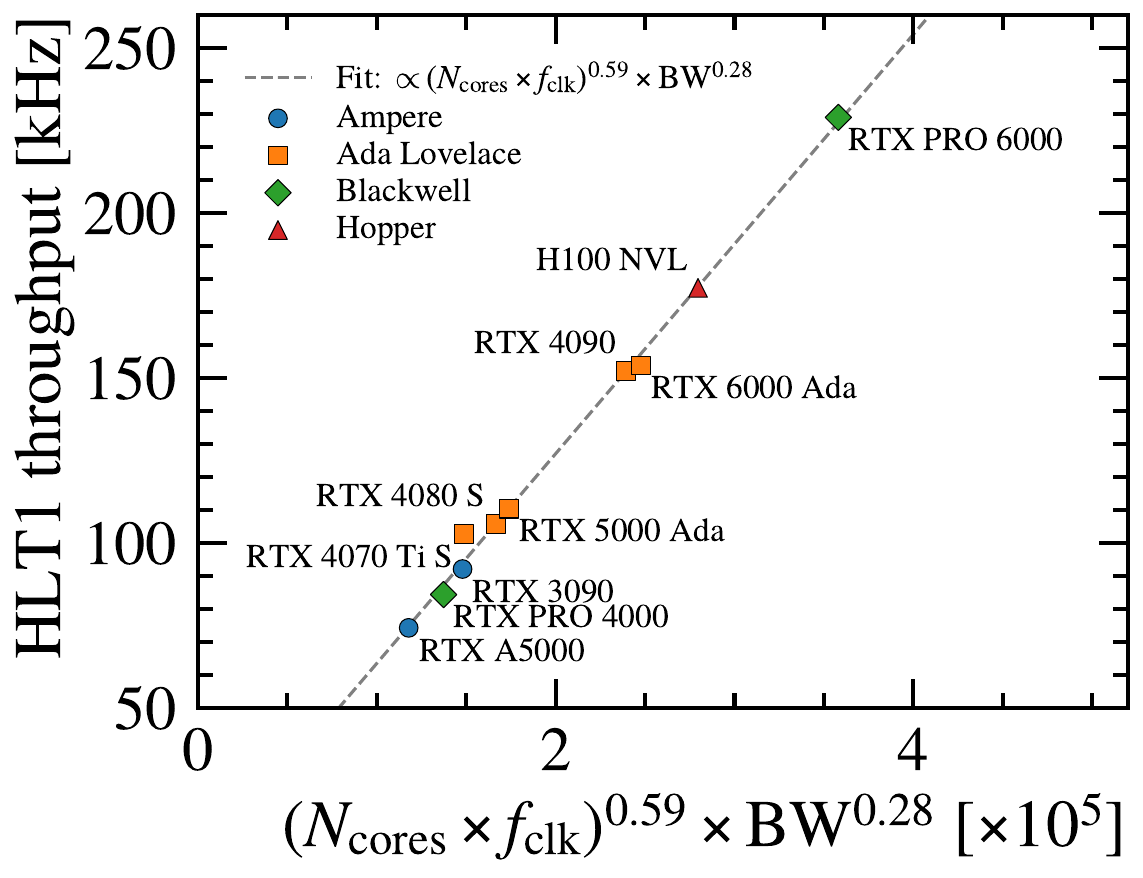}
    \caption{Measured HLT1 throughput versus the fitted model $\text{TP} \propto (N_\text{cores} \times f_\text{clk})^{0.59} \times \text{BW}^{0.28}$ for ten GPUs across four NVIDIA architectures. Each GPU is plotted at its measured hardware parameters. The dashed line shows the fitted model.}
    \label{fig:throughput_model}
\end{figure}

Fitting Eq.~\ref{eq:throughput_model} to the measured throughput of all ten GPUs yields the values
    a$ = 0.59$, b$ = 0.28$ and k$ = 0.64$ of the model parameters.
The result is shown in Fig.~\ref{fig:throughput_model}. All ten GPUs, spanning four NVIDIA architectures and two process nodes, fall on the fitted line with a root-mean-square residual of approximately 3\%.

The results of the fit reveal two properties of the HLT1 workload. The exponent on the compute capacity ($a = 0.59$) is larger than the exponent on the memory bandwidth ($b = 0.28$), which means the throughput depends more on the number of cores and their clock frequency than on the memory bandwidth. This is consistent with the HLT1 reconstruction, which consists of many pattern recognition algorithms with complex control flow rather than simple data-parallel operations. Both exponents are less than one, confirming the sublinear scaling discussed above. In practical terms, doubling the compute capacity increases the throughput by a factor of $2^{0.59} \approx 1.5$, not 2.

Although the model is fitted using measured hardware parameters, the difference between measured and specification values is modest for most GPUs (5--16\% for the SM clock, negligible for the bandwidth). The model can therefore also be applied using specification values to obtain an approximate throughput prediction for GPUs that have not been benchmarked. The accuracy will be lower, because the difference between the specification and the actual clock frequency varies from GPU to GPU. Nevertheless, if the goal is to rank different GPUs by their expected throughput, using specification values should not change the ranking order, because the model is monotonic in both the compute capacity and the bandwidth.

\subsection{Power consumption}

The throughput model from the previous section predicts how fast a given GPU can process HLT1 events. To evaluate the energy efficiency, the power consumption must also be known. One might assume that every GPU reaches its TDP during HLT1 execution. However, as shown in Fig.~\ref{fig:throughput_power_comparison}, only five of the ten GPUs reach their TDP during HLT1 execution. The other five consume 9--28\% less power than their TDP. Using the TDP as the power consumption for all GPUs would overestimate the energy cost of the non-power-limited ones.
A model is therefore developed that can predict, for a given GPU, whether it will be power-limited, and if not, what its actual power consumption will be.

\textbf{Power-limited and non-power-limited GPUs}
\label{sec:throughput_power_classification}

The ten GPUs fall into two groups based on their fabrication process. Both Ampere GPUs (RTX A5000 and RTX 3090) use the Samsung 8\,nm process, and both are power-limited: they reach 99.8\% of their TDP during HLT1 execution. Even the RTX 3090, which has a TDP of 390\,W, is power-limited. This means that the HLT1 workload demands more power than any Ampere GPU can supply, so all Ampere GPUs can be treated as power-limited.

The remaining eight GPUs use the TSMC 4\,nm process. Among these, three are power-limited (RTX 5000 Ada, RTX 6000 Ada, RTX PRO 4000) and five are not (RTX 4070 Ti Super, RTX 4080 Super, RTX 4090, H100 NVL, RTX PRO 6000). Whether a TSMC 4\,nm GPU is power-limited depends on the ratio of its TDP to its core count. The three power-limited TSMC 4\,nm GPUs are all workstation models. NVIDIA designs workstation GPUs for energy efficiency, assigning them a low TDP relative to their core count. Gaming models (RTX 4070 Ti Super, RTX 4080 Super, RTX 4090), by contrast, are designed for maximum performance and have a much higher TDP per core. The same is true for the datacenter H100 NVL and the high-end RTX PRO 6000. These five GPUs all have enough TDP per core to avoid being power-limited.

\textbf{Power demand per core}
\label{sec:throughput_power_demand}

To understand why some GPUs are power-limited and others are not, the power demand of the HLT1 workload itself can be measured. The power demand per core is defined as the measured peak power divided by the number of CUDA cores:
\begin{equation}
    P_\text{core} = \frac{P_\text{measured}}{N_\text{cores}} \,.
    \label{eq:throughput_pcore}
\end{equation}
This is the true power demand of the HLT1 workload per core, because the GPU is not constrained by its TDP. For a power-limited GPU, the measured power cannot exceed the TDP, so $P_\text{core} = \text{TDP} / N_\text{cores}$ is lower than the true demand. The $P_\text{core}$ of power-limited GPUs is therefore biased and cannot be used to measure the workload demand.

\begin{figure}[htb]
    \centering
    \includegraphics[width=0.65\textwidth]{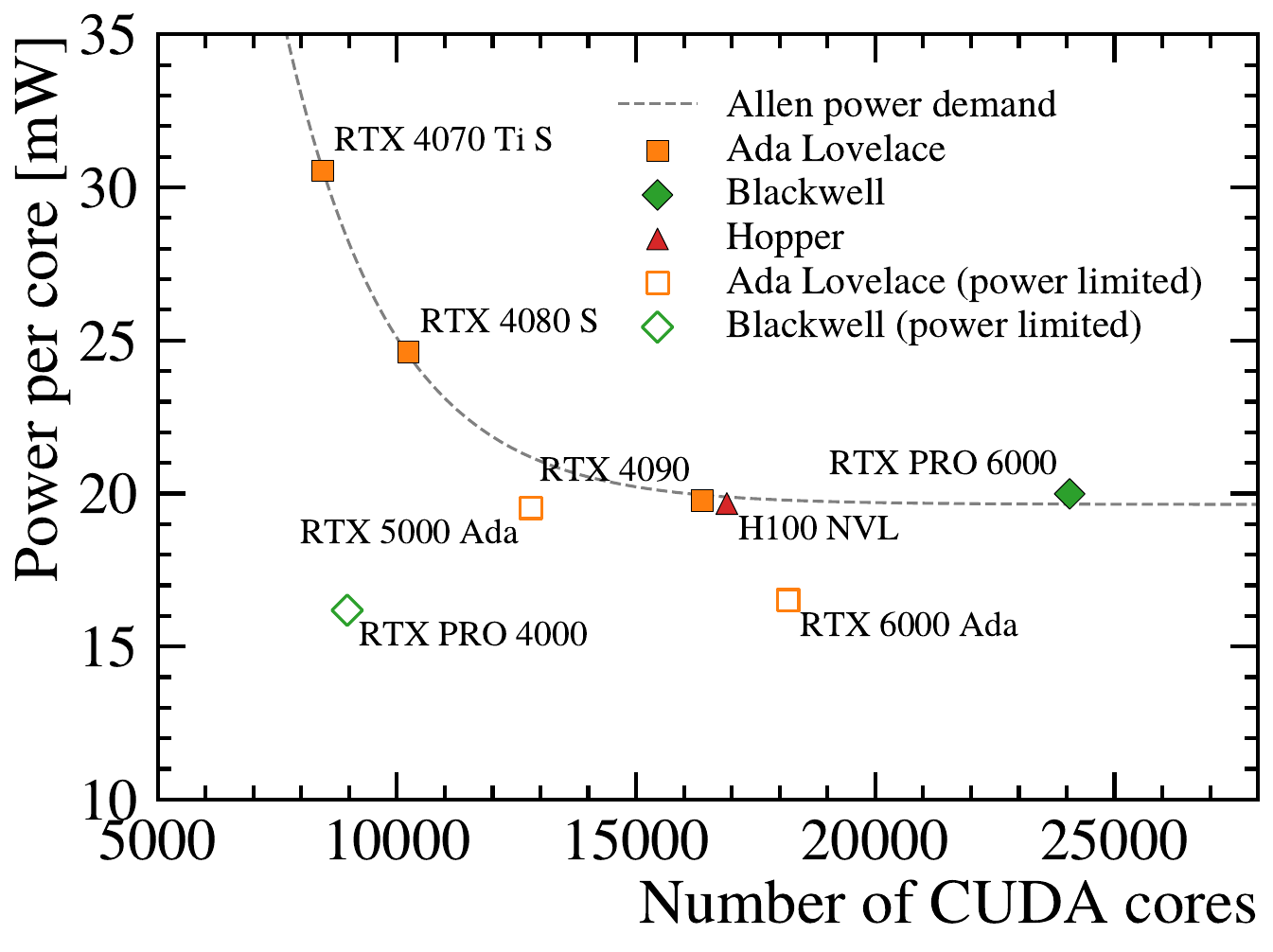}
    \caption{Power per core as a function of the number of CUDA cores for the eight TSMC 4\,nm GPUs. Filled markers show non-power-limited GPUs, open markers show power-limited GPUs. The dashed curve is the fitted power demand model in Eq.~(\ref{eq:throughput_power_demand}). Power-limited GPUs fall below the demand curve, confirming that their TDP is too low to supply the power that HLT1 demands.}
    \label{fig:throughput_power_per_core}
\end{figure}

Fig.~\ref{fig:throughput_power_per_core} shows $P_\text{core}$ as a function of the number of CUDA cores for the eight TSMC 4\,nm GPUs. The Ampere GPUs are excluded because their Samsung 8\,nm fabrication process consumes more power per transistor than the TSMC 4\,nm process, so mixing the two would not produce a consistent curve. In addition, both Ampere GPUs are power-limited, so neither provides an unbiased measurement of the workload demand. The five non-power-limited GPUs (filled markers) show a clear pattern: $P_\text{core}$ decreases with increasing core count and approaches a floor. This trend is described by an exponential decay:
\begin{equation}
    P_\text{core}(N_\text{cores}) = P_\text{floor} + A \times \exp\left(-\frac{N_\text{cores}}{\tau}\right) \,,
    \label{eq:throughput_power_demand}
\end{equation}
with fitted parameters $P_\text{floor} = 19.6$\,mW, $A = 489$\,mW and $\tau = 2223$ cores. This curve represents the power that HLT1 demands from each core as a function of core count on the TSMC 4\,nm process.

The decrease of $P_\text{core}$ with core count reflects how the HLT1 workload is distributed across cores. With fewer cores, each core handles a larger share of the work, so it consumes more power. As the core count increases, the work is spread more thinly and each core consumes less. At very high core counts, however, the HLT1 parallelisation reaches a limit: each core already handles the smallest unit of work that cannot be subdivided further. Beyond this point, adding more cores does not reduce the work per core, and $P_\text{core}$ converges to the floor of approximately 19.6\,mW. This floor is the per-core power cost of HLT1 when the parallelisation is fully exploited on the TSMC 4\,nm process.

\textbf{Power-limiting criterion}
\label{sec:throughput_power_criterion}

The power demand curve serves as a boundary between power-limited and non-power-limited operation. If a GPU's TDP per core ($\text{TDP} / N_\text{cores}$) falls below the demand curve at its core count, the GPU cannot supply enough power per core for HLT1. It will reach its TDP, and the SM clock will be reduced to stay within the power budget. If the TDP per core is above the demand curve, the GPU has enough power and will not be power-limited.

The three power-limited TSMC 4\,nm GPUs (open markers in Fig.~ \ref{fig:throughput_power_per_core}) confirm this criterion. All three fall below the demand curve: their TDP per core values are 19.5\,mW (RTX 5000 Ada), 16.2\,mW (RTX PRO 4000) and 16.5\,mW (RTX 6000 Ada), all below the demand at their respective core counts.

The two Ampere GPUs are not included in the TSMC 4\,nm fit because their fabrication process is different. Both are power-limited, and their measured $P_\text{core}$ values (28.0\,mW for the RTX A5000 and 37.1\,mW for the RTX 3090) are higher than the TSMC 4\,nm demand curve at the same core counts. This is expected, because the older Samsung 8\,nm process uses more power per transistor than the TSMC 4\,nm process. Using the power demand curve, the total power consumption of a TSMC 4\,nm GPU running HLT1 can be predicted as
\begin{equation}
    P_\text{predicted} = \min\left(P_\text{core}(N_\text{cores}) \times N_\text{cores},\; \text{TDP}\right) \,.
    \label{eq:throughput_predicted_power}
\end{equation}
If the demand exceeds the TDP, the GPU is power-limited and the power consumption equals the TDP. Otherwise, the power consumption is given by the demand curve. For Ampere GPUs, the power consumption is taken to be the TDP, since all Ampere GPUs are power-limited for the LHCb HLT1 application.

\subsection{Energy efficiency}
In the previous two sections we developed a model for the HLT1 throughput and a model for the GPU power consumption. Together, they provide all the ingredients needed to predict the energy efficiency of the LHCb HLT1 on any GPU device. The energy efficiency is defined as the number of events processed per joule:
\begin{equation}
    E_\text{eff} = \frac{\text{TP}}{P} \,,
    \label{eq:throughput_energy_efficiency}
\end{equation}
where TP is the throughput and $P$ is the power consumption. For the predicted energy efficiency, the throughput is computed from the throughput model (see Eq.~\ref{eq:throughput_model}) using the specification hardware parameters, and the power consumption is computed from the power demand model in Eq.\ref{eq:throughput_predicted_power}). This allows the energy efficiency to be estimated for any GPU device using only its specification parameters, without running a benchmark.

\begin{figure}[htb]
    \centering
    \includegraphics[width=0.45\textwidth]{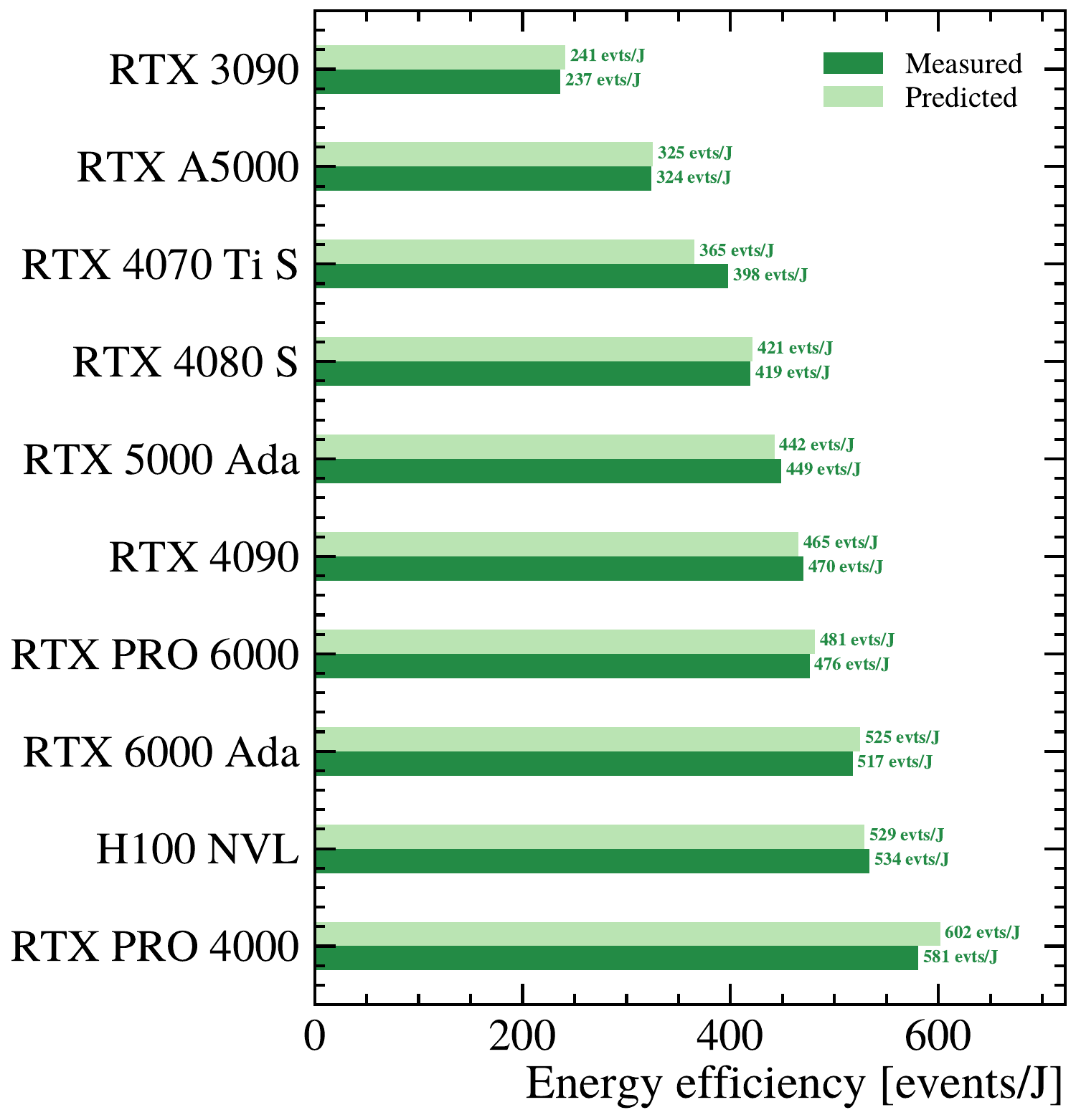}
    \caption{Energy efficiency (events per joule) for each GPU, comparing measured and predicted values. The predicted throughput is obtained from the throughput model in Eq.~\ref{eq:throughput_model} using the GPU specification parameters. The predicted power is obtained from the power demand model (see \ref{eq:throughput_predicted_power}). The GPUs are sorted by measured energy efficiency.}
    \label{fig:throughput_energy_efficiency}
\end{figure}

Figure~\ref{fig:throughput_energy_efficiency} shows the measured and predicted energy efficiency for all ten GPUs, sorted from least to most efficient. The predicted values agree with the measured values within a few percent for all GPUs, and the ranking order is preserved. This validates the combination of the two models and confirms that specification parameters are sufficient for ranking GPUs by energy efficiency. The RTX PRO 4000 Blackwell is the most energy-efficient GPU (581\,events/J), followed by the H100 NVL (534\,events/J) and the RTX 6000 Ada (517\,events/J). The two Ampere GPUs are the least efficient (237--324\,events/J), because the older Samsung 8\,nm process consumes more power per core.

As it is shown Fig.~\ref{fig:throughput_energy_efficiency} the GPU with the highest throughput is not the most energy-efficient. The RTX PRO 6000 Blackwell has the highest throughput (229\,kHz) but ranks only fourth in energy efficiency (476\,events/J), because its power consumption (481\,W) is also high. The RTX PRO 4000 Blackwell, on the other hand, has a modest throughput (84\,kHz) but is the most energy-efficient because its TDP is only 145\,W. Throughput and energy efficiency must therefore be evaluated together when selecting GPUs for any GPU-based processing system.

\section{Discussion}
An important finding of this study is that HLT1 does not reach the TDP on all GPUs, even though the GPU utilisation is at 100\% during the entire benchmark. All streaming multiprocessors are active, but the actual power draw stays below the TDP for five of the ten GPUs. This is because the HLT1 reconstruction involves many pattern recognition algorithms with complex branching and frequent synchronisation between threads. A simpler workload, such as matrix multiplication, would keep all functional units busy and draw more power. HLT1 does not, because threads following different branches of the code leave some functional units unused.

The current HLT1 algorithms were developed and optimised on the RTX A5000, which is power-limited. As the GPU farm is upgraded to newer GPUs that are less likely to be power-limited, the optimisation target should change. On a non-power-limited GPU, reducing the branching and improving the thread utilisation would increase both the throughput and the power consumption, but the throughput gain would outweigh the power increase, resulting in better energy efficiency. HLT1 must therefore be optimised for non-power-limited GPUs, not only for power-limited ones like the A5000.

The throughput and power models developed in this work provide a tool for the LHCb collaboration to evaluate future GPU candidates. Given the specification parameters of a new GPU, the models predict the HLT1 throughput, the expected power consumption and the energy efficiency. These predictions can rank candidate GPUs and identify the most promising options before a full benchmark is performed. The models are validated across four NVIDIA architectures and two process nodes, covering the range of GPUs available today. As new architectures appear, the models can be updated by benchmarking a small number of representative GPUs.

For LHCb specifically, these models are relevant to two future hardware decisions. During Long Shutdown~3, the collaboration will consider upgrading the GPU farm for Run~4. The models allow candidate GPUs to be ranked by both throughput and energy efficiency using only their specification parameters, before committing to procurement. Further ahead, the Upgrade~II programme (Run~5) foresees an increase in instantaneous luminosity by a factor of 7.5 compared to Run~3, requiring a corresponding increase in trigger processing capacity. The throughput model provides a basis for projecting how future GPU generations can meet this demand, and the energy efficiency model helps ensure that the power budget of the upgraded farm remains manageable.

Although this study was carried out using the HLT1 at LHCb as the benchmark framework, the methodology itself is not specific to any application. The approach of modeling throughput and power consumption as functions of GPU specification parameters can be applied to any GPU-based application where performance needs to be projected across different hardware generations. The only requirement is a set of benchmark measurements on a few GPUs to fit the model parameters, after which the model can predict the performance on any GPU whose specifications are known.

\section*{Conflict of Interest Statement}

The authors declare that the research was conducted in the absence of any commercial or financial relationships that could be construed as a potential conflict of interest.

\section*{Author Contributions}
This work has been carried out by J. Zhuo and A. Oyanguren, being J. Zhuo the main developer. A. Oyanguren and L. Fiorini are the promotors and principal investigators of the HIGH-LOW project in Valencia. A. Fernández Casani has mounted and checked the hardware system, in collaboration with J. Zhuo and V. Kholoimov. J. Zhuo and A. Oyanguren are the main editors of the paper.

\section*{Funding}

The result of this work is part of the project TED2021-130852B-I00 (\textit{DESIGN OF HIGH PERFORMANCE ALGORITHMS FOR LOW POWER SUSTAINABLE HARDWARE FOR LHC EXPERIMENTS AND THEIR UPGRADES}), financiaded by MCIN/AEI/10.13039/501100011033 and the EU “NextGenerationEU”/PRTR. V.K. thanks the US NSF cooperative agreement OAC-1836650 (IRIS-HEP) and the Simons Foundation.

\section*{Acknowledgments}

We are grateful to our colleagues from the LHCb Simulation Project for enabling the production of the simulated samples that we use in this study.
We thank LHCb’s Real-Time Analysis project, Online project and Software \& Computing project for its support, for many useful discussions, and for reviewing an early draft of this manuscript. In particular the comments and suggestions of N.~Neufeld, F.~Sborzacchi, V.~Gligorov, D.~vom~Bruch, M.~de Cian, B.~Couturier and C.~Marín were very useful in the elaboration of the document. 


\bibliographystyle{Frontiers-Harvard} 
\bibliography{test}

\end{document}